\documentclass[twocolumn,showpacs,preprintnumbers,amsmath,amssymb]{revtex4}

\usepackage{makeidx,graphics,subfigure}
\usepackage{graphicx}
\usepackage{dcolumn}
\usepackage{bm}
\usepackage{amsmath} 
\usepackage{amssymb}
\def\ket#1{|#1\rangle}

\def\Proj#1#2{|#1\rangle\langle#2|}
\def\sign{\mbox{\rm sgn}}

\begin{document}


\title{Trajectory-constrained optimal local time-continuous waveform controls for state
transitions in $N$-level quantum systems}

\author{Ming Zhang$^{1}$\email{zhangming@nudt.edu.cn},  Jiahua Wei$^{1}$, Weiwei Zhou$^{1}$, Hong-Yi Dai$^{2}$, Zairong Xi$^{3}$, S. G. Schirmer$^{4}$}
\affiliation{$^{1}$%
College of Mechatronic Engineering
and Automation, National University of Defense Technology\\
Changsha, Hunan 410073,  People's Republic of China
}%
\affiliation{$^{2}$%
School of Science, National University of Defense Technology\\%
Changsha, Hunan 410073,  People's Republic of China }
\affiliation{$^{3}$%
Key Laboratory of Systems and Control,
Chinese Academy of Sciences, \\ Beijing, 100080,
People's Republic of China}
\affiliation{$^{4}$%
Department of Applied Mathematics and Theoretical Physics,
University of Cambridge, Wilberforce Road, CB3 0WA, UK
}%

\date{\today}

\begin{abstract}
Based on a parametrization of pure quantum states we explicitly
construct a sequence of (at most) $4N-5$ local time-continuous
waveforms controls to achieve the specified state transition for
$N$-level quantum system when sufficient controls of the
 Hamiltonian are available.  The control magnitudes are further
optimized in terms of the time-energy performance which is the
generalization of the time performance, and then the
trajectory-constrained optimal local time-continuous waveforms
controls including both local sine-waveforms and
$n$-order-polynomial-function-waveform controls are obtained in
terms of time-energy performance.  It is demonstrated that
constrained optimal local $n$st-order-polynomial-function-waveform
controls approach constrained optimal bang-bang controls when
$n\rightarrow\infty$.

\end{abstract}

\maketitle

\section{Introduction}

Control of quantum systems has been recognized as an important issue for
some time with early beginnings of the application of control theory in
the quantum domain dating back to the 1980s~\cite{Huang2,Ong3,Clark4}.
Generally, the main goal of control theory is to find controls leading
the objects to a desired situation. There are usually two ways of
specifying ``a desired prescribed situation'': the controllability
viewpoint and the optimization viewpoint\cite{FZreview}.  The
controllability of quantum systems has been investigated by many
researchers (see Ref.~\cite{Huang2,qcb,dp}). Optimal control theory has
also been successfully applied to the design of open-loop coherent
control strategies in physical chemistry~\cite{d33,d34}.  Recently,
time-optimal control problems for spin systems have been solved to
achieve specified control objectives in minimum
time~\cite{d35,d36,d37}. In general, optimal control problems can only be
solved by using numerical optimization techniques.

When sufficient controls of the Hamiltonian are available, the third
kind of mixed approach may be adopted: one can first construct some
simple controls along a chosen trajectory to achieve a desired state
transition, and then make full use of degree of freedom to optimize
some kind of performances including minimum time performance.  In
this way, trajectory-constrained simple optimal controls are
obtained.  Present technology justifies this third approach for some
super-conducting systems~\cite{PRL96,PRL98,PRA79} and nuclear
magnetic resonance systems(NMR)~\cite{JMO}, for example.  It has
been demonstrated that simple waveforms such as local square wave
function (Bang Bang control) ~\cite{JPA} can be constructed to
achieve the desired state transition. Quite recently, Bang Bang
control has been successfully applied in some physical
systems\cite{Du,Biercuk,Lange,Damodarakurup,Bylander}.  In practice
the control functions need not be restricted to bang-bang controls
but other time-continuous function waveforms such as
triangle-waveforms and sine-waveform may be used to manipulate
quantum states of $N-$level quantum systems. In this paper, the
third kind of mixed viewpoint is adopted and it is demonstrated that
local time-continuous function waveforms including both local
$n$-order-polynomial waveforms and sine waveforms can be constructed
to achieve the desired state transitions.

\section{Prerequisite}

The state of an $N$-level quantum mechanical system is presented by a
vector in $N$-dimensional Hilbert space $H$. In quantum mechanics, the
state $\psi$ is denoted as $|\psi\rangle$ and called \emph{ket}.  To any
$|\psi\rangle$ is associated a linear operator
$\langle\psi|: H\rightarrow{C}$, which is called \emph{bra}. Given a
\emph{ket} $|\psi\rangle$ and a \emph{bra} $\langle\phi|$, we define a
linear operator $|\psi\rangle\langle\phi|:H\rightarrow{H}$.  The state
of an $N$-level quantum system can be expressed as
$|\psi\rangle=\sum_{n=0}^{N-1}{c_{n}}|n\rangle$ with regard to a chosen
basis $|n\rangle\ (n=0,1,2,\ldots,N-1)$ in the Hilbert space.  The
coefficients $c_n$ are a set of complex numbers satisfying the
normalization constraint $\sum^{N-1}_{n=0}|c_{n}|^{2}=1$. By ignoring
the global phase, the coefficients can be expressed as
\begin{equation}
\label{2a}
\begin{pmatrix} c_{0}\\ \vdots\\c_{N-2}\\ c_{N-1} \end{pmatrix}
= \begin{pmatrix}
  \cos\frac{\theta_{1}}{2}\\ \vdots\\
  e^{i\phi_{N-2}}\sin\frac{\theta_{1}}{2}\ldots\sin\frac{\theta_{N-2}}{2}\cos\frac{\theta_{N-1}}{2}\\
e^{i\phi_{N-1}}\sin\frac{\theta_{1}}{2}\ldots\sin\frac{\theta_{N-2}}{2}\sin\frac{\theta_{N-1}}{2}
\end{pmatrix}
\end{equation}
with $0\leq\theta_{1},\ldots,\theta_{N-1}\leq\pi$ and
$0\leq\phi_{1},\ldots,\phi_{N-1}<2\pi$.  Thus any pure state
$\ket{\psi}$ of an $N-$level quantum system can be represented by the
$2(N-1)$ generalized geometric parameters $({\Theta};{\Phi})$ with
$\Theta=(\theta_{1},\ldots,\theta_{N-1})^T$ and
$\Phi=(\phi_{1},\ldots,\phi_{N-1})^T$.

For $N\geq2$ we define the operators ($N\times N$ matrices)
\begin{subequations}
\label{P1}
\begin{align}
  x_{N,k} &= \Proj{k}{k+1}+\Proj{k+1}{k}\\
  y_{N,k} &= i[\Proj{k+1}{k}-\Proj{k}{k+1}]\\
  z_{N,k} &= I_N -2\Proj{k+1}{k+1} \\
  I_{N,k} &= \Proj{k}{k} +\Proj{k+1}{k+1}.
\end{align}
\end{subequations}
for $k=0,1,...,N-2$ where $I_{N}=\sum^{N-1}_{j=0} \Proj{j}{j}$ is
the identity.  For $N=2$ these reduce to the standard Pauli
operators $z_{2,0}=\sigma_z$, $y_{2,0}=\sigma_y$ and
$x_{2,0}=\sigma_x$.

\textbf{Lemma}: Let $0\leq{t_{0}}<t_{1}<\infty$ and
$t\in(t_{0},t_{1})\subset{R^{+}}$.  Both $F(t)$ and $f(t)$ are scalar
time functions defined on $R$.  If $\frac{dF(t)}{dt}=f(t)$ for
${t\in(t_{0},t_{1})}$ and $H$ is a constant Hamiltonian, then the
solution of the operator differential equation
\begin{equation}
  i\dot{X}(t) =  f(t) H X(t)
\end{equation}
 is given by${X}(t)= e^{-i [F(t)-F(t_{0})] H} X(t_0)$ with the initial state ${X}(t_{0})$.

Setting $\Delta F(t)=F(t)-F(t_0)$ evaluation of the matrix exponential
for $H=z_{N,k}$ and $H=y_{N,k}$ yields the explicit formulas
\begin{subequations}
\begin{align}
e^{-i\Delta F(t) z_{N,k}} =
   & e^{-i\Delta F(t)} \{I_{N}+[e^{i 2\Delta F(t)}-1]\Proj{k+1}{k+1} \}
   \label{statez2}\\
e^{-i \Delta F(t) y_{N,k}} =
   & I_{N,k} \cos \Delta F(t) -i y_{N,k} \sin \Delta F(t) \nonumber\\
   & + I_{N}-I_{N,k} . \label{statey2}
\end{align}
\end{subequations}
This Lemma implies that one  have degree of freedom to construct control Hamiltonian to
steer the quantum system to the target state.  In this paper we focus on controls
given by both a local sine waveform
\begin{equation}
\label{fls}
f_{ls}(t;t_{0},t_{1},A)
 =\left\{\begin{array}{cc}
A\sin\frac{\pi\cdot{(t-t_{0})}}{(t_{1}-t_{0})}&t\in[t_{0},t_{1}]\\
0&otherwise
\end{array} \right.
\end{equation}
and a local $n$-order-polynomial-function $f_{ln}(t;t_{0},t_{1},A)$
\begin{equation}
\label{triangle} f_{ln}(t;t_{0},t_{1},A) =\left\{\begin{array}{ll}
-A[\frac{t_{1}+t_{0}-2t}{t_{1}-t_{0}}]^{n}+A& t\in[t_{0},\frac{t_{0}+t_{1}}{2})\\
-A[\frac{2t-(t_{1}+t_{0})}{t_{1}-t_{0}}]^{n}+A& t\in[\frac{t_{0}+t_{1}}{2},t_{1})\\
 0                                     & otherwise
\end{array} \right.
\end{equation}
We further have
$\int^{t_{1}}_{t_{0}}f_{ls}(t;t_{0},t_{1},A)dt=\frac{2A(t_{1}-t_{0})}{\pi}$,
$\int^{t_{1}}_{t_{0}}f_{ln}(t;t_{0},t_{1},A)dt=\frac{{A(t_{1}-t_{0})}n}{n+1}$,
$\int^{\infty}_{0}|f_{ls}(t;t_{0},t_{1},A)|^{2}dt=\frac{A^{2}(t_{1}-t_{0})}{2}$
and
$\int^{\infty}_{0}|f_{ln}(t;t_{0},t_{1},A)|^{2}dt=\frac{A^{2}(t_{1}-t_{0})2n^{2}}{2n^{2}+3n+1}$.

When some waveform controls are chosen to achieve a state transition at
the target time $t_{f}$, we still have some degree of freedom to
optimize the control magnitudes with regard to a chosen performance
index.  Here we shall consider two kind of performance indices, the
transition time $J_{t}=\int_{0}^{t_{f}}{1}dt=t_{f} $ and a combined
time-energy performance index
\begin{equation}
 \label{Jte}
  J_{te} = {t_{f}}+\lambda^{-1}\cdot\int_{0}^{t_{f}}E(u(t))dt %
         = \int_{0}^{t_{f}}[1+\lambda^{-1}E(u(t))]dt
\end{equation}
that takes into account the energy cost $E(u(t))$ of the control
vector $u(t)$ as well as the time required. For example, if
$H(t)=\sum_{i}u_{i}(t)H_{i}$, then $u(t)=(u_{i}(t))$ and
$E(u(t))=\sum_{i}|u_{i}(t)|^{2}$. Here $\lambda>0$ is introduced as
a ratio parameter that defines the relative weight of the energy and
time resource costs, and the equivalent physical unit of $\lambda$
is
$W=J\cdot{s^{-1}}=N\cdot{m}\cdot{s^{-1}}=(kg)\cdot{m^{2}}\cdot{s^{-3}}$.
It should be emphasized that $J_{te}$ is reduced to $J_{t}$ if
$\lambda\rightarrow{\infty}$, so the time-energy performance
$J_{te}$ can be interrupted as the generalization of time
performance.

\section{Trajectory-constrained optimal local sine-waveform transition controls}

Consider an $N-$level quantum system which is governed by the
Schr\"odinger equation (in units of $\hbar=1$)
\begin{equation}
  \label{SE}
  i\tfrac{d}{dt} \ket{\psi(t)} = \sum^{N-2}_{k=0}[{u_{y,k}(t)y_{N,k}}+u_{z,k}(t)z_{N,k}]\ket{\psi(t)}
\end{equation}
where $y_{N,k}$ and $z_{N,k}$ are defined in Eq.~(\ref{P1}).  Such a
drift-free Hamiltonian can be obtained for many systems by transforming
to a rotating frame, suitably expanding the control fields and making
certain simplifying assumptions as discussed e.g.  in Ref. \cite{JPA}.

It has been demonstrated in Ref.~\cite{JPA} that one can steer the system
(\ref{SE}) from an arbitrary initial state $|\psi_{0}\rangle$ to a
target state $|\psi_{s}\rangle$ by using (at most) $4N-5$ Bang-Bang
control based on the parametrization of the initial and target states in
terms of the $2(N-1)$ geometric parameters $(\Theta_{0};\Phi_{0})
=(\theta^{0}_{1},\ldots,\theta^{0}_{N-1};\phi^{0}_{1},\ldots,\phi^{0}_{N-1})^T$
and $(\Theta_{s};\Phi_{s})
=(\theta^{s}_{1},\ldots,\theta^{s}_{N-1};\phi^{s}_{1},\ldots,\phi^{s}_{N-1})^T$
described in Section 2.  One can also achieve the same state transition
by using (at most) $4N-5$ local sine waveform controls and the design
process can be divided into three stages: (1) Steer the quNit from
$(\Theta_{0};\Phi_{0})$ to $(\Theta_{0};\mathbf{0})$ by performing
$(N-1)$ local $Z$-rotations; (2) Transfer the quNit from
$(\Theta_{0};\mathbf{0})$ to $(\Theta_{S};\mathbf{0})$ by performing
$2N-3$ local $Y$-rotations; (3) Manipulate the quNit from
$(\Theta_{s};\mathbf{0})$ to $(\Theta_{s};\Phi_{s})$ by performing
$(N-1)$ local $Z$-rotations.

Let $m=2,3,...,N-1$ and $j=1,2,...,N-1$, we have
\begin{equation}
\label{cy}
\begin{split}
u_{y,0}(t)
 &=\sign(\theta_{1}^{0}-\theta_{1}^{s})f_{ls}(t;t_{2N-3},t_{2N-2},A_{\theta_{1}^{0s}})\\
u_{y,m-1}(t)
 &= f_{ls}(t;t_{2N-2-m},t_{2N-1-m},A_{\theta_{m}^{0}})y_{N,m-1} \\
 &-f_{ls}(t;t_{2N-4+m},t_{2N-3+m},A_{\theta_{m}^{s}}) y_{N,m-1}\\
   u_{z,j-1}(t)
 &=
 \sign(\pi-{\phi_{j}^{0}})f_{ls}(t;t_{j-1},t_{j},A_{\phi_{j}^{0}})\\
 &=\sign(\phi_{j}^{s}-\pi)f_{ls}(t;t_{3N-5+j},t_{3N-4+j},A_{\phi_{j}^{s}})\\
\end{split}
\end{equation}
with control magnitudes $A_{\theta_{m}^{0}}, A_{\theta_{m}^{s}}$,
$A_{\theta_{1}^{0s}}$, $A_{\phi_{j}^{0}}$, and $A_{\phi_{j}^{s}}$
 corresponding to the geometric parameter $\theta_{m}^{0}$,
${\theta_{m}^{s}}$, $\theta_{1}^{0}-\theta_{1}^{s}$, $\phi_{j}^{0}$
and  $\phi_{j}^{s}$.
$t_{j}-t_{j-1}=\frac{\min\{\phi_{j}^{0},2\pi-\phi_{j}^{0}\}\pi}{4A_{\phi_{j}^{0}}}$
$t_{2N-1-m}-t_{2N-2-m}=\frac{\theta_{m}^{0}\cdot{\pi}}{4A_{\theta_{m}^{0}}}$,
$t_{2N-2}-t_{2N-3}=\frac{|\theta_{1}^{0}-\theta_{1}^{s}|\cdot{\pi}}{4A_{\theta_{1}^{0s}}}$,
$t_{2N-3+j}-t_{2N-4+j}=\frac{\theta_{j}^{s}\cdot{\pi}}{4A_{\theta_{j}^{s}}}$
and
$t_{3N-4+j}-t_{3N-5+j}=\frac{\min\{2\pi-\phi_{j}^{s},\phi_{j}^{s}\}\pi}{4A_{\phi_{j}^{s}}}$

Both the transition time $t_{f}=t_{4N-5}-t_{0}$ and the corresponding
energy cost can be expressed in terms of the initial and target state
parameters as well as the magnitudes of sine waveform functions. This
implies that one can optimize the magnitudes of sine wave function in
terms of time-energy performance index given by Eq.~(\ref{Jte}).

Suppose the control amplitudes are bounded by $L$, and denote
$L_{s}^{*}=min(L,\sqrt{2\lambda})$ and
$w_{ls}(x)=\frac{\pi}{4}(\tfrac{1}{x} + \tfrac{x}{2\lambda})$, the
optimal amplitudes for the sine waveform controls, minimizing the
performance index $J_{te}$, are
$A_{\phi_{j}^{0}}=A_{\phi_{j}^{s}}=A_{\theta_{m}^{0}}=
A_{\theta_{m}^{s}}=A_{\theta_{1}^{0s}}=L_{s}^{*}$ and we have
\begin{equation} \label{uJ2} J^{*}_{te}
={t_{f}}+\lambda^{-1}\cdot\int_{0}^{t_{f}}E(u(t))dt=
(C_{1}+C_{2})w_{ls}(L_{s}^{*})
\end{equation}
where $C_1$ and $C_2$ are given by
\begin{equation}
\label{const}
\begin{split}
 C_1 &=\sum^{N-1}_{l=2}(\theta_{l}^{0}+\theta_{l}^{s})+|\theta_{1}^{0}-\theta_{1}^{s}|  \\
 C_2 &=\sum^{N-1}_{k=1}[\min(2\pi-\phi_{k}^{0},\phi_{k}^{0})+\min(2\pi-\phi_{k}^{s},\phi_{k}^{s})].
\end{split}
\end{equation}
The ``best'' time performance for bounded controls is obtained by
letting $\lambda\rightarrow+\infty$.  If $L\rightarrow+\infty$, then time
performance index $J_t\rightarrow0$.

\begin{figure}[ht]
\centering
\subfigure[$2$-rotation  trajectories]
 {\label{a}
\scalebox{0.35}{\includegraphics{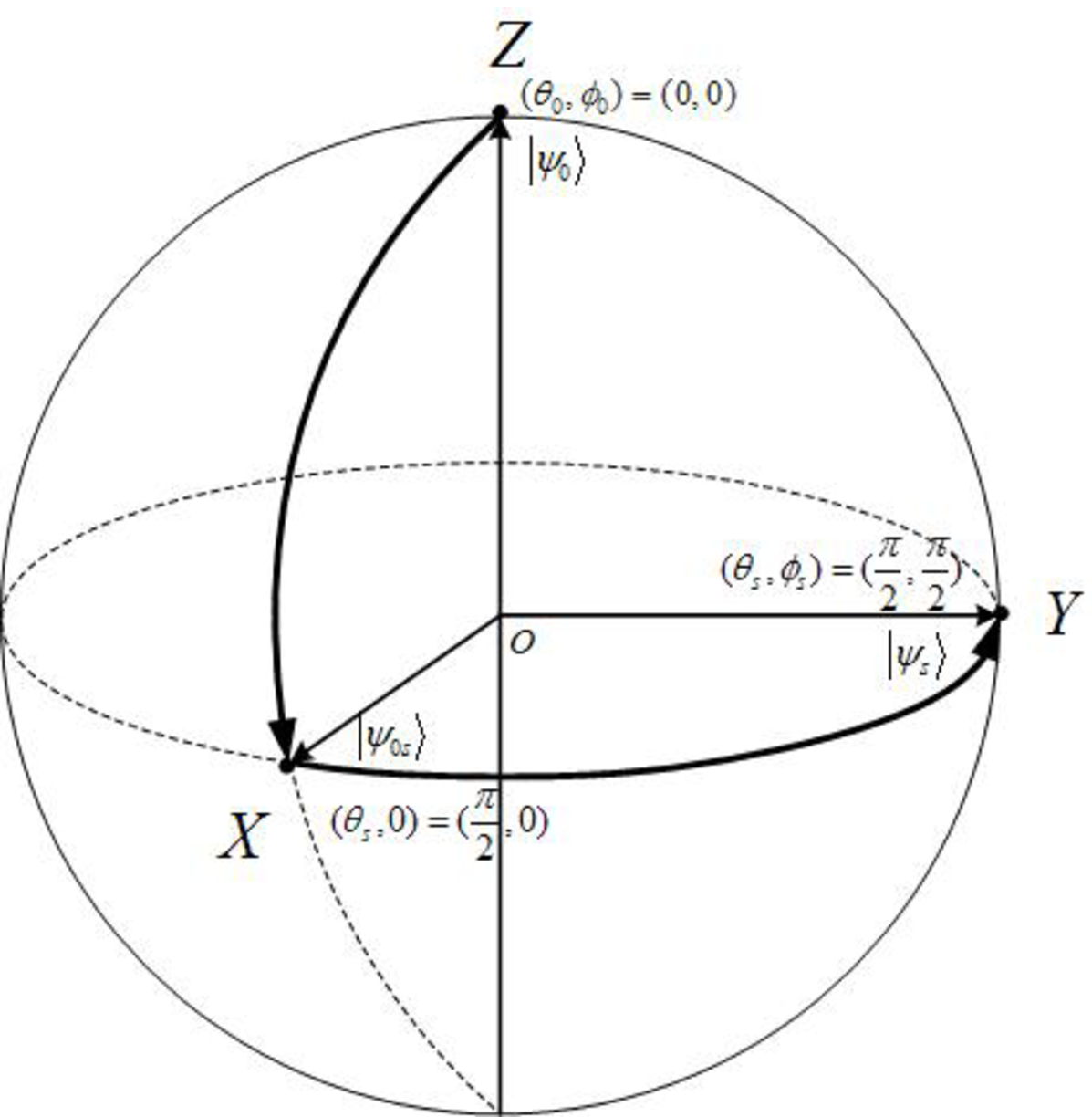}}
}%
\subfigure[$1$-rotation trajectory]
 {\label{b}
\scalebox{0.35}{\includegraphics{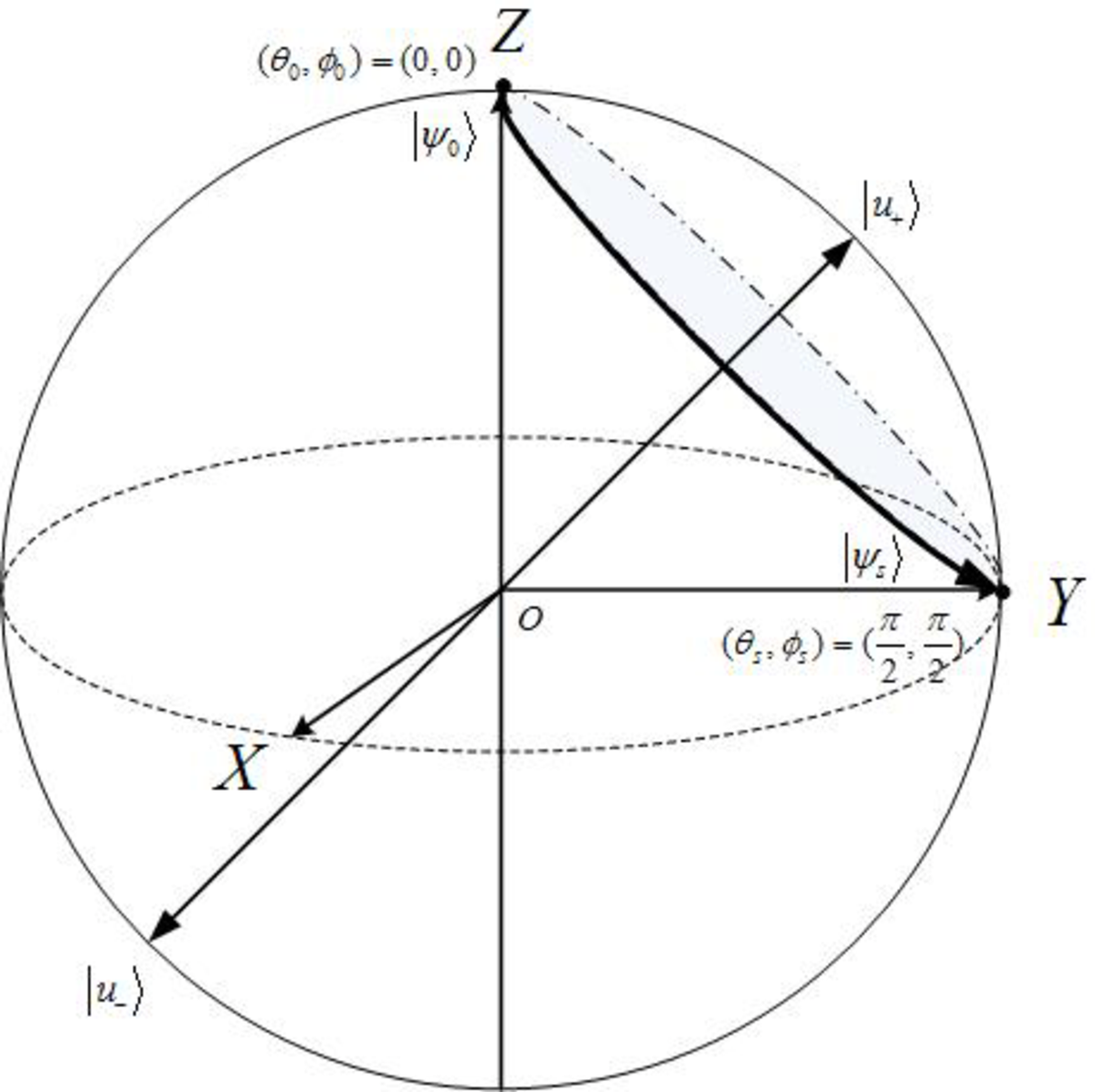}}
}%
\caption{\label{fig1} Two kinds of trajectories on the Bloch sphere}
\end{figure}

To intuitively understand the aforementioned analysis, we further
consider a concrete example by setting $N=2$, $\ket{\psi_{0}}=\ket{0}$,
$\ket{\psi_{s}}=\frac{\sqrt{2}}{2}\ket{0}+i\frac{\sqrt{2}}{2}\ket{1}$,
$\lambda=2$ and $L=1$.  For this concrete example,
$J^{*}_{te}=\frac{5\pi^{2}}{16}$ and the corresponding constrained
optimal bounded sine-waveform controls are given as follow
\begin{equation}
\label{u}
\begin{split}
u^{*}_{z}(t)
=\left\{\begin{array}{ll}
 -\sin\frac{8t-\pi^{2}}{\pi}& t\in[\frac{\pi^{2}}{8},\frac{\pi^{2}}{4})\\
 0                                     & otherwise
\end{array} \right.\\
 u^{*}_{y}(t)=\left\{\begin{array}{ll}
  \sin\frac{8t}{\pi}& t\in[0,\frac{\pi^{2}}{8})\\
 0                                                    & otherwise
\end{array}\right.
\end{split}
\end{equation}
This implies that the desired state transition is achieved by
$2$-rotations as shown in Fig.\ref{fig1}(a). One can also achieve
the same state transition along one-rotation trajectory
(Fig.\ref{fig1}(b)) by choosing
\begin{equation}
\label{1uyz}
 u^{*}_{z}(t)=u^{*}_{y}(t)=\left\{\begin{array}{ll}
  \sin\frac{4\sqrt{2}t}{\pi}& t\in[0,\frac{\sqrt{2}\pi^{2}}{8})\\
 0                                                    & otherwise
\end{array}\right.
\end{equation}
and the corresponding time-energy performance is
$J^{*}_{te}=\frac{3\sqrt{2}\pi^{2}}{16}$.  Therefore, it is exemplified
that the obtained optimal sine-waveform controls proposed in this
section are not globally optimal because they are subject to constraints
on the control waveforms and trajectories.  It should be also underlined
in the aforementioned example that one can generate the target state
$\ket{\psi_{s}}=\frac{\sqrt{2}}{2}\ket{0}+i\frac{\sqrt{2}}{2}\ket{1}$ at
$t_{k}=\frac{\sqrt{2}\pi^{2}}{8}+\frac{k\sqrt{2}\pi^{2}}{4}$(
$k=0,1,2,...$) by performing the sine waveform functions
$u^{*}_{z}(t)=u^{*}_{y}(t)=\sin\frac{4\sqrt{2}t}{\pi}$ on the controlled
qubit.

\section{Discussion and Conclusion}

Following similar analysis as in Section 3, one can obtain
trajectory-constrained optimal $n$st-order polynomial waveform controls.
Denote $w_{lb}(x)=\frac{1}{2}(\tfrac{1}{x} + \tfrac{x}{\lambda})$
$w_{ln}(x)=(\tfrac{n+1}{2nx} + \tfrac{nx}{(2n+1)\lambda})$,
$L^{*}_{B}=\min(L,\sqrt{\lambda})$,
$L^{*}_{n}=\min(L,\frac{\sqrt{(2n+1)(2n+2)\lambda}}{2n})$ and $t_*=(C_1+
C_2)/L$ with constants $C_1$ and $C_2$ defined as Eq.~(\ref{const}).
The time and time-energy performances are further summered in Table
\ref{tab.1} for constrained optimal Bang-Bang (BB) controls, constrained
optimal local sine-waveform (LS) controls, and constrained optimal local
$n$-order-polynomial-function-waveforms(LN) controls.

\begin{table}
\caption{Trajectory-constrained minimum transition times $J_t$ for
bounded controls with amplitude bounds given by $L$  and optimal
time-energy performance $J_{te}$ for both bounded and unbounded
controls for different waveforms. } \label{tab.1}
\begin{center}
\begin{tabular}{lccc}
Case & $J_{t}$  & $J_{te}$(bounded) &  $J_{te}$(unbounded) \\
BB   & $\tfrac{1}{2}t_*$
     & $w_{lb}(L^{*}_{B})(C_{1} +C_{2})$
     & $\sqrt{\lambda}(C_1+C_2)$ \\
LN   & $\tfrac{n+1}{2n}t_*$
     & $w_{ln}(L^{*}_{n})(C_1+C_2)$
     & $\frac{\sqrt{(2n+1)(2n+2)\lambda}}{2n}(C_1+C_2)$  \\
LS   & $\tfrac{\pi}{4} t_*$
     & $ w_{ls}(L_{s}^{*})(C_1+C_2)$
     & $\frac{\pi\sqrt{2\lambda}}{4}(C_1+C_2)$
\end{tabular}
\end{center}
\end{table}

In this technical communique, we utilize the geometric parametrization
of quantum states and the properties of generalized Pauli operators to
develop analytic control schemes and then construct optimal local
time-continuous function controls to achieve state transitions for
multi-level quantum systems.  We use the remaining degrees of freedom to
optimize the magnitude parameters of the time-continuous function with
regard to time-energy performance.  Constrained optimal controls of both
sine-waveform and $n$-order-polynomial-function-waveform are obtained.
When $n\rightarrow{\infty}$, constrained optimal LN controls approach to
the constrained optimal BB controls, generalizing the results in
\cite{JPA}.

The choice of a time-energy performance index given by Eq.~(\ref{Jte})
is motivated by experimental feasibility considerations, which require
that the desire for fast state transition be balanced against the need
to limit the amount of energy required to achieve the transition.  It
should be underlined that Eq.~(\ref{Jte}) is reduced to $J_{t}$ when
$\lambda\rightarrow{\infty}$, therefore time-energy performance can be
interpreted as a generalization of time performance.

Although the trajectory-constrained optimal time-continuous waveform
controls are not globally optimal, the resulting constrained optimal
controls have the advantage of being of a simple form, which can be
given analytically without the burden for numerical optimization.  It is
also exemplified in section 3 that periodic sine waveforms controls could
possibly be utilized to periodically generate the target state, and this
observation implies that bounded control with finite frequencies can be
used for quantum state engineering.

 \emph{Acknowledge} This work was partially supported by the National Nature Science
Foundation of China under Grant Nos. 60974037, 61134008, 11074307,
60774099 and 60821091.  SGS also acknowledges funding from EPSRC via ARF
Grant EP/D07192X/1 and Hitachi.


\end{document}